\documentclass[runningheads]{llncs}
\usepackage[T1]{fontenc}
\usepackage{graphicx}
\usepackage{xcolor}
\usepackage{amsmath}

\begin{document}
\title{Self-Supervised Learning for Physiologically-Based Pharmacokinetic Modeling \\ in Dynamic PET 
}
\titlerunning{Spatio-Temporal Self-Supervised Network for PBPK Modeling}

\author{Francesca De Benetti\inst{1} \and
Walter Simson\inst{2} \and
Magdalini Paschali\inst{3} \and
Hasan Sari\inst{4,5} \and
Axel Romiger \inst{5} \and
Kuangyu Shi\inst{1,5} \and
Nassir Navab\inst{1} \and
Thomas Wendler\inst{1,6}
}
\authorrunning{F. De Benetti et al.}

\institute{Chair for Computer Aided Medical Procedures and Augmented Reality, Technische Universit\"{a}t M\"{u}nchen, Garching, Germany, \email{\{francesca.de-benetti,wendler\}@tum.de}
\and
Department of Radiology, Stanford University School of Medicine, Stanford, USA
\and
Department of Psychiatry and Behavioral Sciences, Stanford University School of Medicine, Stanford, USA
\and
Advanced Clinical Imaging Technology, Siemens Healthcare AG, Lausanne, Switzerland
\and
Department of Nuclear Medicine, Inselspital, Bern University Hospital, Bern, Switzerland
\and
SurgicEye GmbH \& ScintHealth GmbH, Munich, Germany}

\newcommand\asteriskfill{\leavevmode\xleaders\hbox{$\ast\ $}\hfill\kern0pt}

\maketitle 
\begin{abstract}
Dynamic positron emission tomography imaging (dPET) provides temporally resolved images of a tracer enabling a quantitative measure of physiological processes. Voxel-wise physiologically-based pharmacokinetic (PBPK) modeling of the time activity curves (TAC) can provide relevant diagnostic information for clinical workflow. Conventional fitting strategies for TACs are slow and ignore the spatial relation between neighboring voxels. We train a spatio-temporal UNet to estimate the kinetic parameters given TAC from F-18-fluorodeoxyglucose (FDG) dPET. This work introduces a self-supervised loss formulation to enforce the similarity between the measured TAC and those generated with the learned kinetic parameters. 
Our method provides quantitatively comparable results at organ-level to the significantly slower conventional approaches, while generating pixel-wise parametric images which are consistent with expected physiology.
To the best of our knowledge, this is the first self-supervised network that allows voxel-wise computation of kinetic parameters consistent with a non-linear kinetic model.
\end{abstract}

\section{Introduction}
Positron Emission Tomography (PET) is a 3D imaging modality using radiopharmaceuticals, such as F-18-fluorodeoxyglucose (FDG) as tracer. 
Newly introduced long axial field-of-view (LA-FoV) PET scanners have enabled dynamic PET (dPET) with frame duration < 1 minute
\cite{surti2020total}, allowing the observation of dynamic metabolic processes throughout the body. For a given voxel in space, the measurement of radioactivity concentration over time can be described by a characteristic curve, known as Time Activity Curve (TAC), measured in [Bq/ml].

TACs can be described by mathematical functions, called physiologically-based pharmacokinetic (PBPK) models or kinetic models (KM) \cite{pantel2022principlesI}. The parameters of the KM represent physiologically relevant quantities and are often called \textit{micro-parameters}, whereas their combinations are called \textit{macro-parameters}~\cite{pantel2022principlesI,watabe_kinetic_2016}. While the former can be retrieved only by methods that directly use the KM function, the latter can be computed also by simplified linearized methods (such as the Logan and the Patlak-Gjedde plots).

The approaches to extract KM parameters can be split in two categories: Volume of Interest (VoI) methods, in which the average TAC in a VoI is used, or voxel-based methods.
Despite the former displaying less noise and therefore lower variance in the kinetic parameters, VoI-based methods only provide organ-wise information. On the other hand, voxel-based methods allow the generation of parametric images, in which kinetic parameters are visualized at a voxel level, but suffer from motion and breathing artifacts~\cite{watabe_kinetic_2016}.

Parametric images are reported to be superior in lesion detection and delineation when compared to standard-of-care static activity- and weight-normalized PET images, known as Standard Uptake Value (SUV) images~\cite{dimitrakopoulou2021kinetic,fahrni2019does}.
Biomarkers in breast cancer (e.g. recurrence rate and mortality risk) can be extracted by FDG-dPET kinetic analysis. Changes in the kinetic parameters during oncological therapy are associated with pathological response to treatment, whereas this is not true for changes in SUV~\cite{pantel2022principlesII}.

Despite the advantages of parametric images in diagnosis, the generation of accurate micro-parametric images is not possible in clinical practice today.
To address the problem of the calculation of micro-parameteric images, we propose the use of a custom 3D UNet~\cite{cciccek20163d} to estimate kinetic \textit{micro-parameters} in an unsupervised setting drawing inspiration from physics informed neural networks (PINN). The main contributions of this work are:
\begin{itemize}
    \item A self-supervised formulation of the problem of kinetic \textit{micro-parameters} estimation
    \item A spatio-temporal deep neural network for parametric images estimation
    \item A quantitative and qualitative comparison with conventional methods for PBPK modeling
\end{itemize}

\subsection{Related work}

Finding the parameters of a KM is a classical optimization problem~\cite{besson202018f,wang2022total,zuo2019structural} solved by fitting the kinetic model to measured TACs in a least squares sense~\cite{AVULA2003219,pantel2022principlesI,snyman2005practical}.
The non-linearity of the KM functions make this approach prone to overfitting and local minima, and sensitive to noise~\cite{pantel2022principlesI}. Therefore, non-linear parametric imaging is still too noisy for clinical application~\cite{watabe_kinetic_2016}.
To limit the drawbacks of the non-linear parameter fitting, the identification of kinetic parameters is commonly performed using simplified linearized versions of the KM~\cite{dimitrakopoulou2021kinetic,watabe_kinetic_2016}, such as the Patlak-Gjedde plot~\cite{dias2022normal,sari2022first}, which are often included in clinical software for KM such as PMOD\footnote{https://www.pmod.com}.

Preliminary works towards KM parameter estimation in dPET imaging have recently begun to be explored.
Simulated data were used by Moradi et al. to select models to describe kinetic data~\cite{moradi2022fdg}. Using an auto-encoder along with a Gaussian process regression block, they estimated KM parameters using a simulated 1D input function. 
Li et al. presented a similar approach for the prediction of the radioisotope uptake rate parameter for quantification of myocardial blood flow from simulated PET sinograms~\cite{li2022direct}.
Huang et al. used a supervised 3D U-Net with residual modules to predict a macro-parameter image using an SUV image derived from static PET as input, and a Patlak image derived from dPET acquisition as ground truth~\cite{huang2022parametric}. 
Cui et al. proposed a conditional deep image prior framework to predict a macro-parametic image using a DNN in an unsupervised setting~\cite{cui2022unsupervised}. Until now, methods used simulated data~\cite{li2022direct,moradi2022fdg} or static PET~\cite{huang2022parametric}, were supervised~\cite{li2022direct,moradi2022fdg,huang2022parametric} or predicted \textit{macro-parameters}~\cite{huang2022parametric,cui2022unsupervised}.

\begin{figure}[t]
    \centering
    \includegraphics[width=\textwidth]{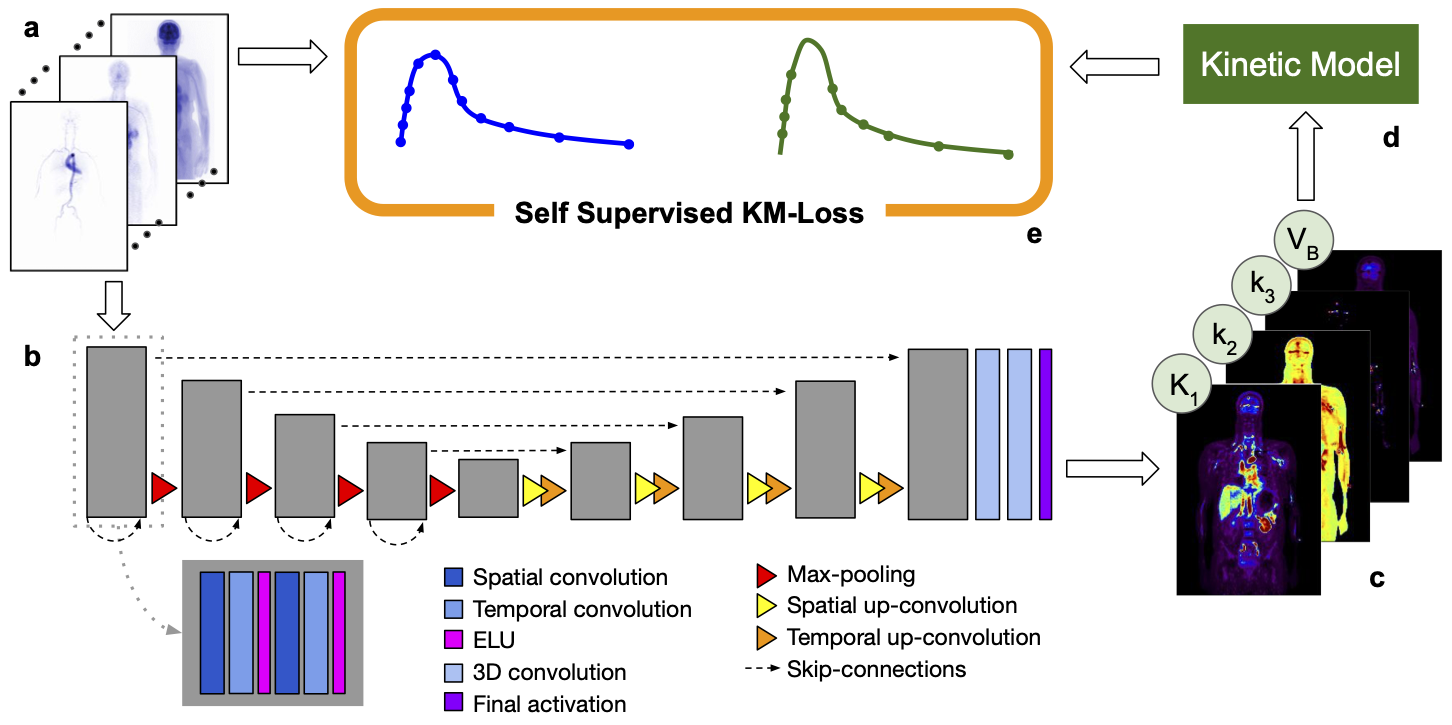}
    \caption{Pipeline for the proposed approach: dPET images (a) are processed by a UNet (b) featuring spatial and temporal convolutions to estimate the parameters of the kinetic model (d), here with 4 parametric images (c). For training, the input TACs are compared with the TACs inferred by the UNet using mean-square error (e).}
    \label{fig:pipeline}
\end{figure}

\section{Methodology and Materials}
We propose to compute the kinetic \textit{micro-parameters} in a self-supervised setting by directly including the KM in the loss function and comparing the modeled to the measured TAC.  For this reason, an understanding of the KM is fundamental to describe our pipeline (see Fig. \ref{fig:pipeline}). We will briefly introduce the KM and then the proposed method in this Section.
\subsection{Kinetic modelling}
An FDG-TAC can be described by an irreversible two-compartment kinetic model (2TC), which represents the interaction of two compartments, $F(t)$ (free) and  $B(t)$ (bound), and takes as input the radioactivity concentration in blood or plasma $A(t)$~\cite{pantel2022principlesI}. The concentration of the tracer $\tilde{C}(t)$ in each tissue can be described as a set of ordinary differential equations~\cite{watabe_kinetic_2016}:
\begin{align} \label{eq:2TC}
    \frac{d F(t)}{dt} &= K_1 A(t) - (k_2+k_3)F(t)\nonumber \\
    \frac{d B(t)}{dt} &= k_3 F(t)\\
    \tilde{C}(t) &= F(t) + B(t)  \nonumber 
\end{align}
where the \textit{micro-parameters} $K_1$ [ml/cm$^3$/min], $k_2$ [1/min], and $k_3$ [1/min] 
represent physiologically meaningful quantities~\cite{dimitrakopoulou2021kinetic,watabe_kinetic_2016}.
Including the blood fraction volume $V_B$
allows to correctly model the contribution to the radioactivity from vessels in a voxel that are too small to be resolved by the PET scanner~\cite{pantel2022principlesI}. 

Together, the TAC of each voxel in a FDG dPET acquisition can be modeled as 
$C(t) = (1-V_B)\tilde{C}(t)+V_BA(t)$, and solved using the Laplace transform~\cite{watabe_kinetic_2016}:
\begin{equation}\label{eq:laplace_2TC}
   C(t) = (1-V_B) \bigg[\frac{K_1}{k_2+k_3}\big[k_3+k_2e^{(k_2+k_3)t}\big]\ast A(t)\bigg] + V_B A(t).
\end{equation}

\subsection{Proposed Pipeline}
Our network takes as input a sequence of 2D axial slices and returns a 4-channel output representing the spatial distribution of KM parameters of a 2TC for FDG metabolisation~\cite{pantel2022principlesI}.
The network has a depth of four, with long~\cite{cciccek20163d} and short skip connections~\cite{kustner2020cinenet}. The kernel size of the max-pooling is [2, 2, 2]. After the last decoder block, two 3D convolutional layers (with kernel size [3, 3, 3] and [64, 1, 1]) estimate the kinetic parameters given the output feature of the network. Inside the network the activation function is ELU and critically batch normalization is omitted. The network was trained with an initial learning rate of $10^{-4}$, which was divided by half every 25 epochs, for a maximum of 500 epochs.

Following the approach taken by K{\"u}stner et al. for motion correction of 4D spatio-temporal CINE MRI, we used spatial-temporal convolutions for KM parameter estimation~\cite{kustner2020cinenet}. For PET, we replaced a conventional 3D convolutional layer with (2+1)D spatial and temporal convolutional layers. The spatial convolutional layer is a 3D convolutional layer with kernel size [1, 3, 3] in [t, x, y]. Similarly, the temporal convolutional layer has kernel size of [3, 1, 1].

We imposed that the parameters predicted by the network satisfy Eq. \ref{eq:laplace_2TC} by including it in the computation of the loss. At a pixel-level, we computed the Mean Squared Error (MSE) between the TAC estimated using the corresponding predicted parameters ($\tilde{\text{TAC}}_i$) and the measured one $\text{TAC}_i$, as seen in Fig. \ref{fig:pipeline}.
 
We introduced a final activation function to limit the output of the network to the valid parameter domain of the KM function. Using the multi-clamp function, each channel of the logits is restricted to the following parameter spaces: $K_1 \in [0.01, 2]$, $k_2 \in [0.01, 3]$, $K_3 \in [0.01, 1]$, and $V_B \in [0, 1]$.
The limits of the ranges were defined based on the meaning of the parameter (as in $V_B$), mathematical requirements (as in the minimum values of $k_2$ and $k_3$, whose sum cannot be zero)~\cite{dimitrakopoulou2021kinetic} or previous knowledge on the dataset derived by the work of Sari et al.~\cite{sari2022first} (as in the maximum values of $K_1$, $k_2$ and $k_3$).

To evaluate the performance of the network, we used the Mean Absolute Error (MAE) and the cosine similarity (CS) between $\text{TAC}_i$ and $\tilde{\text{TAC}}_i$.

\subsection{Curve fit} For comparison, parameter optimization via non-linear fitting was implemented in Python using the \texttt{scipy.optimize.curve\_fit} function (version 1.10), with initial values set to 0.1 for $K_1$ and $k_2$ and to 0.01 for $k_3$ and $V_B$, bounds set to $[0, +\infty)$ and step equal to 0.001.

\subsection{Dataset}
The dataset is composed by 23 oncological patients with different tumor types (see Supplementary Material). 
dPET data was acquired on a Biograph Vision Quadra
for 65 minutes, over 62 frames. The exposure duration of the frames were 2$\times$10 s, 30$\times$2 s, 4$\times$10 s, 8$\times$30 s, 4$\times$60 s, 5$\times$120 s and 9$\times$300 s. The PET volumes were reconstructed with a voxel size of $(1.65 \text{ mm})^3$. 
The dataset included one label map of 8 organs (bones, lungs, heart, liver, kidneys, spleen, aorta, brain) and one Image‐Derived Input Function (IDIF, [Bq/ml]) from the descending aorta (which represented $A(t)$) per patient.

The PET frames and the label map were resampled to isotropic spacing of 2.5 mm. Then, the dataset was split patient-wise into training, validation and test set, with 10, 4 and 9 patients respectively. The training set consisted of 750 slices and the validation consisted of 300. In both cases 75 axial slices per patient were extracted in pre-defined patient-specific range from the the lungs to the bladder (included) and were cropped to size 112$\times$112 pixels.

\begin{table}
\centering
\caption{Demographics.}\label{tab:demo}
\begin{tabular}{|c|c|c|c|c|c|c|}
\hline
ID & Age [years] & Sex & Weight [kg] & Injected dose [MBq] & Diagnosis & Dataset \\
\hline
1 & 68 & Male & 82 & 295.0 & Lung Cancer & Train\\
2 & 75 & Female & 65 & 190.0 & Breast Cancer & Test\\
3 & 46 & Female & 70 & 217.0 & Breast Cancer & Test\\
4 & 70 & Female & 53 & 168.0 & Lung Cancer & Test\\
5 & 39 & Female & 57 & 178.0 & Breast Cancer & Train\\
6 & 62 & Male & 70 & 213.0 & Lymphoma & Test \\
7 & 85 & Male & 78 & 248.7 & Lymphoma & Train \\
8 & 80 & Female & 94 & 274.5 & Fallopian tube Cancer & Train\\
9 & 77 & Male & 81 & 261.0 & Lymphoma & Test \\
10 & 29 & Male & 130 & 398.0 & Lymphoma & Validation \\
11 & 64 & Male & 80 & 247.0 & Gastric Cancer & Train\\
12 & 53 & Male & 68 & 208.0 & Lung Cancer & Validation\\
14 & 59 & Male & 85 & 276.0 & Lymphoma & Validation \\
15 & 40 & Female & 63 & 191.0 & Cervical Cancer & Train \\
16 & 54 & Male & 79 & 236.0 & Lymphoma & Test \\
17 & 66 & Male & 94 & 279.0 & Lung Cancer & Train \\
18 & 68 & Male & 79 & 227.0 & Lymphoma & Train \\
19 & 66 & Male & 81 & 258.0 & Lymphoma & Test \\
20 & 73 & Female & 59 & 182.0 & Breast Cancer & Test\\
21 & 28 & Male & 85 & 252.0 & Lymphoma & Test \\
22 & 65 & Female & 65 & 196.0 & Head \& Neck Cancer & Train \\
23 & 59 & Male & 91 & 262.0 & Melanoma & Test \\
24 & 68 & Male & 80 & 227.0 & Melanoma & Train \\
\hline
Mean & 60.1 $\pm$ 15.3& 15/9 & 76.6 $\pm$ 16.7 & 235.3 $\pm$ 51.3 & - & -\\
\hline
\end{tabular}
\end{table}

\section{Results}
Tab. \ref{tab:ablation} shows the results of the 8 ablation studies we performed to find the best model. We evaluated the impact of the design of the convolutional and max-pooling kernels, as well as the choice of the final activation.
The design of the max pooling kernel (i.e., kernel size equal to [2, 2, 2] or [1, 2, 2]) had no measurable effects in terms of CS in most of the experiments, with the exception of Exp. 3.2 (as seen in Tab. \ref{tab:ablation}) where max-pooling only in space resulted in a drop of 0.06. When evaluating the MAE, the use of the 3D max-pooling was generally better. 
The most important design choice is the selection of the final activation function. Indeed, the multi-clamp activation function was proven to be the best both in terms of CS (Exp 4.1: CS = 0.78  $\pm$ 0.05) and MAE (Exp 4.2: MAE = 3.27 $\pm$ 2.01). 
Compared to the other final activation functions, when the multi-clamp is used the impact of the max-pooling design is negligible also in terms of MAE. For the rest of the experiments, the selected configuration is the one from Exp. 4.1 (see Tab. \ref{tab:ablation}).

\begin{table}[t]
\caption{Configurations and metrics of the ablation studies for architecture optimization.}\label{tab:ablation}
\centering
\begin{tabular}{|c|c|c|c|c|c|c|c|c|}
\hline
Exp & Convolution & Pooling & Final & Activation & Batch & CS $\uparrow$ & MAE $\downarrow$\\
 &  &  & activation & & normalization & & \\
\hline
1.1 & 3D & 3D & Absolute & ELU & No & 0.74 $\pm$ 0.05 & 3.55 $\pm$ 2.12\\
1.2 & 3D & space & Absolute & ELU & No & 0.74 $\pm$ 0.05 & 3.64 $\pm$ 2.21\\
2.1 & space + time & 3D & Absolute & ELU & No & 0.75 $\pm$ 0.05 & 3.59 $\pm$ 2.33\\
2.2 & space + time & space & Absolute & ELU & No & 0.75 $\pm$ 0.05 & 3.67 $\pm$ 2.20\\
3.1 & space + time & 3D & Clamp & ELU & No & 0.75 $\pm$ 0.05 & 3.48 $\pm$ 2.04\\
3.2 & space + time & space & Clamp & ELU & No & 0.69 $\pm$ 0.05 & 3.55 $\pm$ 2.07\\
4.1 & space + time & 3D & Multi-clamp & ELU & No & \textbf{0.78  $\pm$ 0.05} & 3.28 $\pm$ 2.03 \\
4.2 & space + time & space & Multi-clamp & ELU & No & 0.77 $\pm$ 0.05 & \textbf{3.27 $\pm$ 2.01}\\
- & space + time & 3D & Multi-clamp & ReLU & No & 0.70 $\pm$ 0.04 & 4.05 $\pm$ 2.37\\
- & space + time & space & Multi-clamp & ReLU & No & 0.69 $\pm$ 0.04 & 4.05 $\pm$ 2.33\\
- & space + time & 3D & Multi-clamp & ELU & Yes & 0.57 $\pm$ 0.03 & 5.31 $\pm$ 2.20\\
- & space + time & space & Multi-clamp & ELU & Yes & 0.62 $\pm$ 0.03 & 4.82 $\pm$ 2.31\\
\hline
\end{tabular}
\end{table}

\begin{figure}[b]
    \centering
    \includegraphics[width=\textwidth]{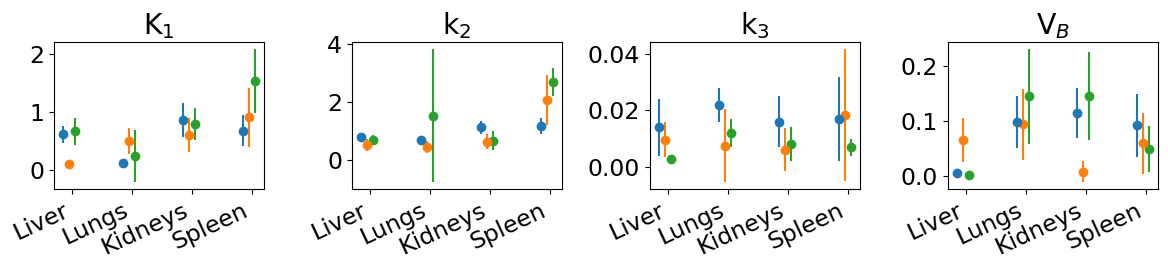}
    \caption{Comparison between the kinetic parameters obtained with the DNN (proposed) \textbf{(blue)}, and curve fit on the 9 test set patients \textbf{(orange)}. For comparison of value statistics, a curve fit \textbf{(green)} on all 23 patients~\cite{sari2022first} is also provided.}
    \label{fig:kinetic_organ}
\end{figure}

Fig. \ref{fig:kinetic_organ} shows the kinetic parameters for four selected organs as computed with the proposed DNN (P$_{\text{DNN}}$), and as computed with curve fit (P$^{\text{ref}}_{\text{CF}}$) using all 23 patients~\cite{sari2022first} and using only the test set (P$_{\text{CF}}$). The voxel-wise kinetic parameters predicted by the DNN were averaged over the available organ masks. For parameter estimation with real data there is no ground truth, so this comparison serves as plausibility check.

In terms of run-time, the DNN needed approx. one minute to predict the kinetic parameters of the a whole-body scan ($\approx$ 400 slices), whereas curve fit took 8.7 minutes for a single slice: the time reduction of the DNN is expected to be $\approx$ 3.500 times.

\begin{table}
\centering
\caption{P$_{\text{DNN}}$: Organ-level kinetic parameters as predicted by the best model (Exp 4.1) in the test set (9 patients), including the organs in Figure 2 of the submission plus bones, aorta and heart. These values are computed by averaging the voxel-wise \textit{micro-parameters} in each VoI.}
\begin{tabular}{|c|c|c|c|c|}
\hline
VOI & $K_1$ [ml/cm$^3$/min]& $k_2$ [1/min]& $k_3$ [1/min]& $V_B$ [$\cdot$]\\
\hline
Liver & 0.611 $\pm$ 0.146 & 0.793 $\pm$ 0.135 & 0.014 $\pm$ 0.010 & 0.005 $\pm$ 0.006\\
Lungs & 0.116 $\pm$ 0.039 & 0.683 $\pm$ 0.060 & 0.022 $\pm$ 0.006 & 0.098 $\pm$ 0.048\\
Kidneys & 0.867 $\pm$ 0.292 & 1.135 $\pm$ 0.213 & 0.016 $\pm$ 0.009 & 0.115 $\pm$ 0.046 \\
Spleen & 0.678 $\pm$ 0.270 & 1.165 $\pm$ 0.277 & 0.017 $\pm$ 0.015 & 0.092 $\pm$ 0.057 \\
Bones & 0.112 $\pm$ 0.039 & 0.619 $\pm$ 0.037 & 0.025 $\pm$ 0.005 & 0.005 $\pm$ 0.004 \\
Aorta & 0.657 $\pm$ 0.475 & 1.346 $\pm$ 0.777 & 0.035 $\pm$ 0.028 & 0.622 $\pm$ 0.238\\
Heart & 0.644 $\pm$ 0.159 & 1.108 $\pm$ 0.257 & 0.020 $\pm$ 0.007 & 0.376 $\pm$ 0.133\\
\hline
\end{tabular}
\end{table}

\begin{table}
\centering
\caption{P$_{\text{CF}}$: Organ-level kinetic parameters as computed with the curve fit method on the test set (9 patients), including the organs in Figure 2 of the submission plus bones, aorta and heart. These values are computed by averaging the TAC in each VoI and then applying the curve fit method to those.}
\begin{tabular}{|c|c|c|c|c|}
\hline
VOI & $K_1$ [ml/cm$^3$/min]& $k_2$ [1/min]& $k_3$ [1/min]& $V_B$ [$\cdot$]\\
\hline
Liver & 0.598 $\pm$ 0.294 & 0.643 $\pm$ 0.256 & 0.006 $\pm$ 0.008 & 0.007 $\pm$ 0.020\\
Lungs & 0.102 $\pm$ 0.031 & 0.539 $\pm$ 0.206 & 0.010 $\pm$ 0.006 & 0.065 $\pm$ 0.040\\
Kidneys & 0.503 $\pm$ 0.221 & 0.448 $\pm$ 0.199 & 0.008 $\pm$ 0.013 & 0.094 $\pm$ 0.066 \\
Spleen & 0.904 $\pm$ 0.509 & 2.074 $\pm$ 0.872 & 0.018 $\pm$ 0.023 & 0.059 $\pm$ 0.056 \\
Bones & 0.063 $\pm$ 0.031 & 0.332 $\pm$ 0.178 & 0.018 $\pm$ 0.010 & 0.014 $\pm$ 0.021 \\
Aorta & 0.220 $\pm$ 0.155 & 0.322 $\pm$ 0.223 & 0.008 $\pm$ 0.005 & 0.608 $\pm$ 0.226\\
Heart &0.434 $\pm$ 0.143 & 0.569 $\pm$ 0.215 & 0.011 $\pm$ 0.007 & 0.318 $\pm$ 0.108\\
\hline
\end{tabular}
\end{table}

\section{Discussion}
In the architecture of the network, the selection of the kernel design is important. Even though the choice of the final activation function has greater impact, using spatial and temporal convolution results in an increase in the performance (+0.01 in CS) and reduces the number of trainable parameters (from 2.1 M to 8.6 K), as pointed out by~\cite{kustner2020cinenet}. Therefore, the convergence is reached faster. Moreover, the use of two separate kernels in time and space is especially important because voxel counts are influenced by the neighbouring voxels due to the limited resolution of the PET scanner~\cite{watabe_kinetic_2016}, but they are not influenced by the previous/following acquisitions in time.
In general, there is good agreement between P$_{\text{DNN}}$,  P$_{\text{CF}}$ and P$^{\text{ref}}_{\text{CF}}$. The DNN prediction of $K_1$ in the spleen, $k_2$ in the kidneys and in the spleen, and $k_3$ in the lungs is outside of the confidence interval of the P$^{\text{ref}}_{\text{CF}}$ results published by Sari et al.~\cite{sari2022first}. 

\begin{figure}[t]
    \centering
    \includegraphics[width=\textwidth]{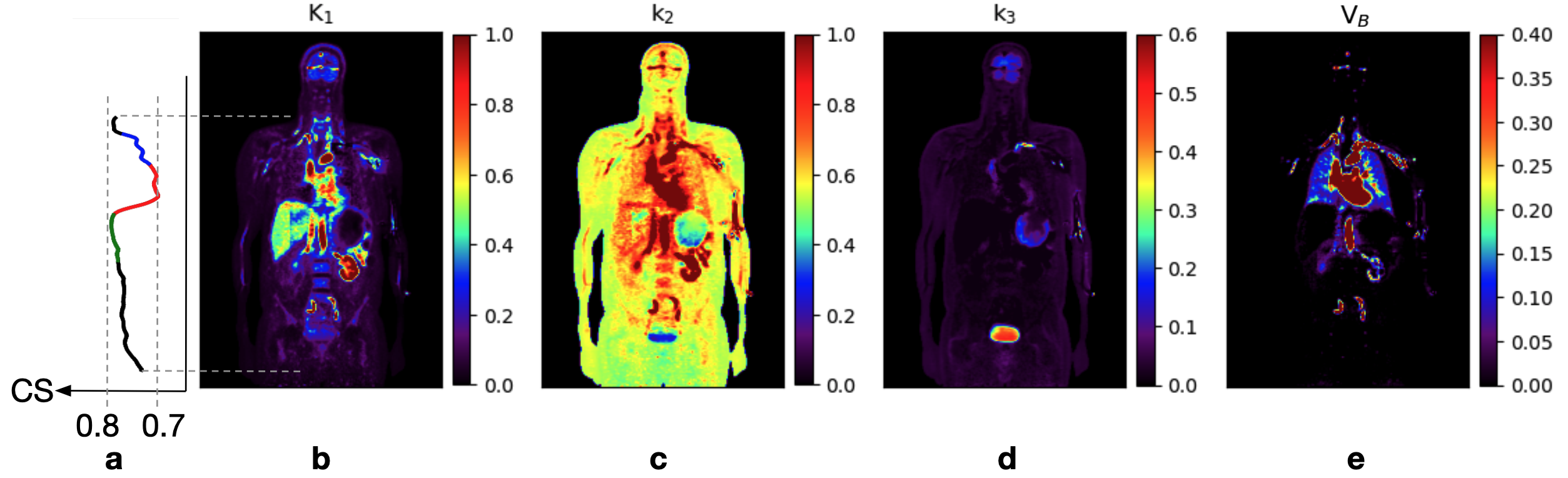}
    \caption{(a) Cosine similarity (CS) per slice in patient 23 (blue: lungs; red: lungs and heart; green: liver). (b-e) Parametric images of a coronal slice for the same patient.}
    \label{fig:param_img_28}
\end{figure}

An analysis per slice of the metrics shows that the CS between $\text{TAC}_i$ and $\tilde{\text{TAC}}_i$ changes substantially depending on the region: CS$_\text{max}$=0.87 within the liver boundaries and 
CS$_\text{min}$ =0.71 in the region corresponding to the heart and lungs (see Fig. \ref{fig:param_img_28}a). This can be explained by the fact that $V_B$ is underestimated for the heart and aorta. The proposed network predicts $V^{\text{heart}}_B = 0.376 \pm  0.133$ and $V^{\text{aorta}}_B = 0.622 \pm  0.238$ while values of nearly 1 are to be expected. This is likely due to breathing and heart beat motion artifacts, which cannot be modeled properly with a 2TC KM that assumes no motion between frames.

\begin{figure}[t]
    \centering
    \includegraphics[width=0.8\textwidth]{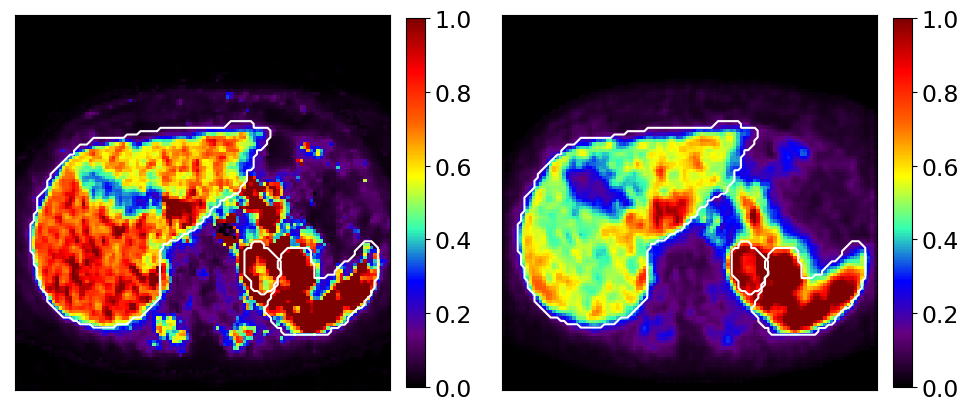}
    \caption{Comparison of $K_1$ parametric images for an axial slice through the liver of patient 2 in [ml/cm$^3$/min] (left: curve fit, right: DNN). The contours visible are the liver (left), the spleen (center) and the left kidney (right).}
    \label{fig:comparison}
\end{figure}

Fig. \ref{fig:param_img_28}b-e show the central coronal slice of the four parametric images through an exemplary patient. All parametric images show expected trends: e.g., $K_1$, the uptake rate is high in heart, liver and kidney; similarly, the blood fraction volume $V_B$ is higher in the heart, blood vessels and lungs.
This trend is also visible in Fig. \ref{fig:comparison}a where an axial slice of $K_1$ is depicted. $K_1$ as calculated by the DNN distributes homogeneously when compared to the $K_1$ image calculated using curve fit. Also, the distribution in the liver is more realistic in the DNN image, where the gallbladder can be seen as an ellipsoid between right and left liver lobes. High $K_1$ regions are mainly within the liver, spleen and kidney for the DNN, while they also appear in unexpected areas in the curve fit image (e.g., next to the spine or in the region of the stomach).

Standard non-linear optimization algorithms pitfalls, including local minima and overfitting, can result in poor estimation of kinetic parameters if performed independently~\cite{pantel2022principlesI}. Hence, the calculation of kinetic parameters at voxel level is more difficult than VoI-approaches that are commonly used. Moreover, voxel-wise estimation requires more computational power or simplified methods like the Patlak plot~\cite{watabe_kinetic_2016}. 

The major limitation of this work is the lack of ground truth and a canonical method to evaluate quantitatively its performance. This limitation is inherent to PBPK modeling and results in the need for qualitative analyses based on expected physiological processes. A possible way to leverage this would be to work on simulated data, yet the validity of such evaluations strongly depends on how realistic the underlying simulation models are.
As seen in Fig. \ref{fig:param_img_28}a, motion (gross, respiratory or cardiac) has a major impact on the quality of the estimation. Registering different dPET frames has been shown to improve conventional PBPK models~\cite{guo2022mcp} and would possibly have a positive impact on our approach.

\section{Conclusion} 

In this work, inspired by PINNs, we combine a self-supervised spatio-temporal DNN with a new loss formulation considering physiology to perform kinetic modeling of FDG dPET. We compare the best DNN model with the most commonly used conventional PBPK method, curve fit. While no ground truth is available, the proposed method provides similar results to curve fit, but qualitatively more plausible images in terms of physiology and with a radically shorter run-time. Further, our approach can be easily extended to more complex KM 
without increasing the complexity and the need of computational power significantly. Overall this work offers scalability and a new direction of research for analysing pharmacokinetics.

\bibliographystyle{splncs04}
\bibliography{bibliography.bib}

\end{document}